# Multi-slice passband bSSFP fMRI at ultra-high field

Olivier Reynaud, Analina R. da Silva, Rolf Gruetter and Ileana O. Jelescu[1]

Centre d'Imagerie Biomédicale, Ecole Polytechnique Fédérale de Lausanne, Lausanne, Switzerland

**Abstract**

Balanced steady-state free precession (bSSFP) can be used as an alternative to gradient-echo (GE) EPI for BOLD functional MRI when image distortions and signal drop-outs are severe such as at ultra-high field. However, 3D-bSSFP acquisitions have distinct drawbacks on either human or animal MR systems. On clinical scanners, 3D imaging is suboptimal for localized fMRI applications, and also results in distortions, blurring, and increased sensitivity to motion or physiological noise. On pre-clinical systems, 3D acquisitions have low temporal resolution due to limited acceleration options, while single slice offers insufficient coverage. The aim of the present study was to implement a multi-slice bSSFP acquisition with Cartesian read-out to obtain non-distorted BOLD fMRI activation maps in the human and rat brain at ultra-high field. We show that the bSSFP signal characteristics are preserved in a new pseudo-steady-state. In the human brain at 7 Tesla, we demonstrate that both task- and resting-state fMRI can be performed with 2D-bSSFP, with a temporal SNR that matches that of 3D-bSSFP, resulting in – at least – equal performance. In the rat brain at 14 Tesla, we show that the multi-slice bSSFP protocol has similar sensitivity to gradient-echo EPI for task fMRI, while benefitting from much reduced distortions and drop-outs. The advantages of passband bSSFP at 14 Tesla in comparison with GE-EPI are expected to be even more marked for mouse fMRI.

**Abbreviations**: BOLD, blood oxygenation level dependent; bSSFP, balanced steady-state free precession; EPI, echo planar imaging; fMRI, functional magnetic resonance imaging; GE, gradient echo; SNR, signal-to-noise ratio.

---

[1] Address all correspondence to: ileana.jelescu@epfl.ch

## 1. Introduction

The most widespread imaging sequence used in functional MRI (fMRI) is gradient-echo echo planar imaging (GE-EPI). This sequence offers good sensitivity to the blood oxygenation level dependent (BOLD) signal via its $T_2^*$ contrast, while allowing a high temporal resolution. However, at ultra-high field, GE-EPI suffers from two significant drawbacks related to $B_0$ inhomogeneity: spatial distortions making registration to an anatomical reference difficult, and, more importantly, significant signal loss in areas of interest due to susceptibility mismatches at air – tissue interfaces.

On the other hand, it has been reported that balanced steady-state free precession (bSSFP) can measure a significant BOLD effect (Bowen et al., 2005) while offering two key advantages at ultra-high field: it has the highest signal-to-noise ratio (SNR) efficiency of all known pulse sequences, and exhibits minimal susceptibility-related distortions and signal loss (Scheffler and Lehnhardt, 2003). Initially, bSSFP fMRI experiments were performed in the transition-band regime, where indeed dramatic signal magnitude and phase changes were observed with activation (Miller et al., 2003; Scheffler et al., 2001). However, this method is demanding in terms of shimming and offers limited brain coverage for a given phase cycling angle. Passband bSSFP does not suffer from these drawbacks and has also been shown to display excellent BOLD sensitivity, comparable to that of GE-EPI (Bowen et al., 2005) and superior to TR-matched Cartesian GE (Miller et al., 2007; Park et al., 2011).

The sources of BOLD contrast in passband bSSFP are multiple and their relative weight depends on sequence parameters and field strength. At short TE/TR, the BOLD contrast is largely due to $T_2$ changes, while increasing the TE/TR will introduce more $T_2^*$ weighting, similar to GE (Kim et al., 2012; Miller, 2012; Miller et al., 2007; Zhong et al., 2007). While most spins will remain within the broad passband of the bSSFP profile during rest and activation, some contribution to signal change from frequency variations (similar to those that dominate in transition band bSSFP) can also contribute (Miller and Jezzard, 2008), especially at ultra-high field where intra-voxel frequency distributions are broader.

These sources of contrast also impact the spatial specificity of the bSSFP BOLD signal. A short TR increases the relative contribution of extravascular signal around capillaries rather than larger veins and improves specificity. However, a short TR implies that intravascular spins – be they in capillaries or larger veins – will also contribute to the BOLD signal. Previous simulation work, validated by experiments at 3 Tesla, calculated that the intravascular contribution represented indeed a large fraction of the bSSFP fMRI contrast, but that high field and short TR have the potential to provide a localization of the activation closer to the capillaries (Kim et al., 2012). Another study at 9.4 T in rats (Park et al., 2011) reported a spatial shift in the activation foci with different phase cycling angles in the bSSFP acquisitions, which implied that spins contributing to the BOLD contrast may stem from different pools depending on the spatial location and on the magnetic field inhomogeneities.

The requirement to maintain a steady-state implies the use of bSSFP as a 3D technique. However, the need for high temporal resolution in fMRI combines poorly with Cartesian 3D acquisitions, particularly in rodent imaging where acceleration options are very limited. Aside from a recent implementation of compressed sensing for bSSFP fMRI in the rat at 9.4T (Han et al., 2015), rodent studies so far have been limited to single slice acquisitions (Cheng et al., 2014; Muir and Duong, 2011; Park et al., 2011; Zhou et al., 2012). The latter preclude sufficient coverage to effectively describe entire functional regions or large-scale resting-state networks.

On clinical MR systems, 3D bSSFP is typically accelerated using radial (Benkert et al., 2016), spiral (Lee et al., 2008), or multi-line read-outs (Ehses and Scheffler, 2017), as well as parallel imaging (Chappell et al., 2011). However, 3D acquisitions can be sub-optimal for specific applications such as brainstem or retina fMRI, due to increased sensitivity to motion and physiological noise (Miller et al., 2007) and blurring along the second phase-encode direction.

Therefore the aim of the present study was to implement a multi-slice passband bSSFP in the transient state for BOLD fMRI at ultra-high field (7 Tesla for human experiments and 14 Tesla for rat experiments). The use of bSSFP in the transient state has been proposed previously for X-nuclei applications (Scheffler, 2003). Here, we first characterize the transient state through simulations and confront experimental signal profiles with simulated ones. We then compare the performance of 2D multi-slice bSSFP to that of 3D bSSFP in human studies, and to that of single-slice bSSFP and multi-slice GE-EPI in rodent studies of BOLD fMRI.

## 2. Materials and methods
### 2.1. Human brain at 7 Tesla

**Simulations**

Bloch simulations were used to investigate the impact of prolonged relaxation between slab repetitions (number of slices: 1/2/4/8) on the bSSFP magnetization profile. Simulations covered the frequency offset range $\left[-\frac{1}{TR}, \frac{1}{TR}\right]$ and flip angles between 5° and 40°. $T_1$ and $T_2$ values were fixed at human gray matter values at 7 Tesla: 2000 ms and 55 ms, respectively (Wright et al., 2008; Yacoub et al., 2003). The impact of realistic tissue modeling, diffusion and RF pulse duration on spin history (Bieri and Scheffler, 2007, 2009) was neglected. Simulations included the "catalyzation" (also called ramped preparation) module included with the product sequence on the scanner: a series of ten RF pulses with linear flip angle increase (e.g. $-\alpha/4, +\alpha/2, -3\alpha/4, +\alpha, -\alpha$, etc.), applied before each slice acquisition for rapid convergence towards pseudo-steady-state (Deshpande et al., 2003; Le Roux, 2003) – **Figure 1**. The frequency range $\left[-\frac{1}{TR}, \frac{1}{TR}\right]$ was covered using incremental phase cycling steps (Patterson et al., 2011), as implemented at the scanner. bSSFP profile perturbations due to this phase cycling increase were also simulated.

**Experimental**

The study was approved by the Committee for ethics in human research of the canton of Vaud (CER-VD, Switzerland). Experiments were performed on healthy volunteers using a 7T head-only system (Siemens, Erlangen, Germany) and a 32-channel receiver coil (Nova Medical,Inc., MA, USA). Informed written consent was obtained from all participants.

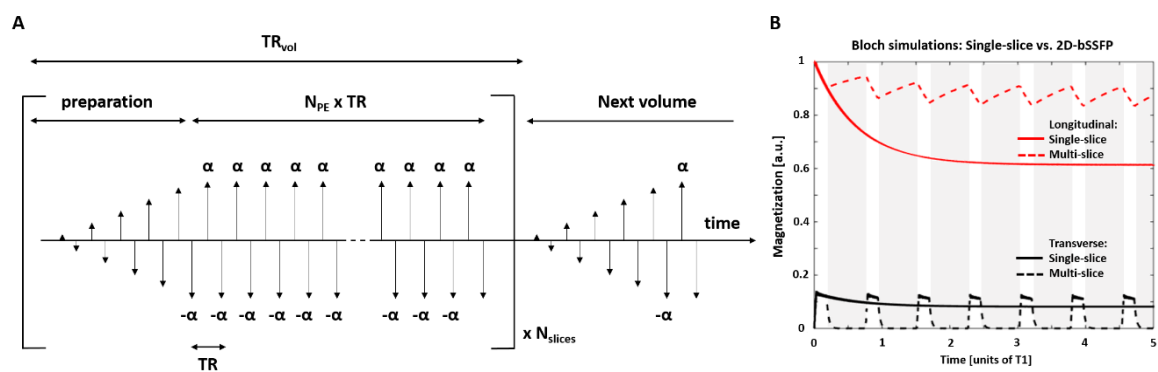

**Figure 1. A**: Chronogram of the multi-slice bSSFP sequence in the passband regime. Ten preparation pulses with linearly increasing flip angle accelerate the convergence towards steady-state. **B**: Evolution of magnetization (transverse – black, longitudinal – red) during single-slice/multi-slice bSSFP (bold/dashed lines). Partial $T_1$ recovery in a given slice is achieved during the acquisition of the other slices (gray area), resulting in a distinct pseudo-steady-state (with higher signal) after a duration ~5$T_1$'s. Only a few transient volumes need be discarded for fMRI.

Single- and multi-slice 2D-bSSFP signal profiles were acquired using: TE/TR=2.9/5.8ms, matrix 52x64, in-plane resolution 3.5x3.5 mm$^2$, slice thickness 4.5 mm, with TR$_{vol}$ = 0.375/1.5 s and FA=15/30° for one and four slices, respectively. The frequency range $\left[-\frac{1}{TR}, \frac{1}{TR}\right]$ was covered using incremental phase cycling steps (Patterson et al., 2011).

To compare single-slice and multi-slice functional contrast, 2D-bSSFP resting-state fMRI (rs-fMRI) data were acquired on 3 subjects (10 min/run) using the parameters above. Resting state networks (RSN) were extracted using the MELODIC toolbox of FSL (Smith et al., 2004) after brain extraction, smoothing (Gaussian kernel, FWHM = 6 mm) and high-pass filtering ($f > 0.01$ Hz).

2D multi-slice (pseudo-steady-state) and 3D (steady-state) bSSFP images with matching parameters were acquired on six subjects: TE/TR = 2.51/5.02 ms, matrix = 64x60x8, in-plane/through-plane resolution = 3.5/4.5mm, optimal FA = 40/15° for 2D/3D, TR$_{vol}$ = 3000/2600 ms, phase-cycling: 180°. The volume TR was longer for 2D due to the ramped preparation before each slice acquisition. 2D and 3D normalized temporal SNR maps (tSNR = $\frac{\bar{s}_t}{\sigma_t \cdot \sqrt{TR_{vol}}}$) were compared at whole-brain level (WB) and in gray matter (GM), white matter (WM) and cerebrospinal fluid (CSF) ROIs in mid-brain.

To compare 2D multi-slice and 3D bSSFP fMRI performance, a finger-tapping task was run on three subjects. Each fMRI run consisted in blocks of 30 s of rest and 30 s of right-hand finger-tapping at 1 Hz frequency, repeated for five minutes. The data were analyzed using SPM12 (http://www.fil.ion.ucl.ac.uk/spm/software/spm12/). The first six volumes were discarded to retain only 2D/3D bSSFP data in pseudo/true steady-state. The pre-processing pipeline included motion correction, slice timing correction and in-plane smoothing (kernel FWHM equal to the voxel size). After general linear model (GLM) analysis using the canonical BOLD hemodynamic response function, the number of activated voxels at $p < 0.05$ after family-wise error correction (FWE), and maximum/average T-scores amongst activated voxels (T$_{max}$/T$_{mean}$) were compared.

### 2.2. Rat brain at 14 Tesla

**Animal preparation**

This study was approved by the Service for Veterinary Affairs of the canton of Vaud. Male Sprague-Dawley rats (Charles River, L'Arbresle, France) (N = 13; weight = 269 ± 18 g) were initially anesthetized with isoflurane (4% for induction and 2% for maintenance) and positioned in a homemade MRI cradle. Respiration rate and rectal temperature were monitored throughout the experiment (SA Instruments, Stony Brook, NY, USA). A catheter was inserted subcutaneously on the back of the animal for medetomidine delivery. For electrical stimulation during the fMRI experiment, two pairs of stainless steel electrodes were inserted in the forepaws, between digits 2 and 3 and digits 4 and 5.

When the animal setup was finalized, the anesthesia was switched from isoflurane to medetomidine (Dorbene, Graeub, Switzerland): an initial bolus of 0.1 mg/kg was injected subcutaneously; the isoflurane was discontinued 10 minutes later (an oxygen/air supply of 20%/80% was maintained throughout the experiment) and a continuous infusion of 0.1 mg/kg/h medetomidine was started another five minutes later. The total sedation time under medetomidine was around three hours, with an average respiration rate of 85 breaths per minute. Medetomidine is free from the vasodilatory effects of isoflurane, which are deleterious both for BOLD sensitivity (Takuwa et al., 2012) and vein segmentation at ultra-high field (Ciobanu et al., 2012).

At the end of the experiment, animals were woken up with an intramuscular injection of atipamezole (Alzane, Graeub, Switzerland) at 0.5 mg/kg and returned to their cages.

**Multi-slice "pseudo steady-state" characterization**

Rat data were acquired on a 14-T Varian system (Abingdon, UK) equipped with 400 mT/m gradients. An in-house built quadrature surface coil was used for transmission and reception.

The aim of the multi-slice (ms) bSSFP was to match the brain coverage, spatial and temporal resolutions typically obtained with GE-EPI on the same system: five 1-mm thick coronal slices (interleaved acquisition), 300 – 400 μm in-plane resolution, and 1.5 s, respectively. Sequence parameters are given in **Table 1**.

|  | Single-slice bSSFP | Multi-slice bSSFP | Multi-slice GE-EPI |
|---|---|---|---|
| TE / TR (ms) | 3.064/ 6.128 | | 17 / 1500 |
| NS | 1 | 5 | 5 |
| DS | 0 | 16 | 0 |
| DV | 53 | 6 | 6 |
| $TR_{vol}$ (s) | 0.2 | 1.5 | 1.5 |
| Flip angle (°) | 12 | 22 | 62 |
| Bandwidth (kHz) | 27 | | 250 |
| Matrix | 64 x 32 | | 64 x 64 |
| In-plane resolution (μm$^2$) | 300 x 400 | | 359 x 359 |
| Slice thickness (mm) | 1 | | 1 |
| Field of view (mm$^2$) | 19.2 x 12.8 | | 23 x 23 |

**Table 1.** Acquisition parameters for bSSFP and GE-EPI protocols. NS: number of slices; DS: dummy scans; DV: dummy volumes; ms: multi-slice; ss: single-slice.

Bloch simulations written in Matlab were used to generate theoretical signals for these multi-slice protocols, using measured relaxation times in the rat cortex at 14 Tesla ($T_2$ = 25 ms / $T_1$ = 2200 ms) and various flip angles. Properties of the transient signal in the multi-slice acquisition were compared to the steady-state signal.

Experimental single-slice and multi-slice bSSFP profiles were measured by varying the phase-cycling angle φ from 0 to 360° in steps of 30°, acquired in random order, with other imaging parameters listed in **Table 1**. The first few volumes were discarded until slab steady-state was established.

**Task-based fMRI**

For each fMRI run, either the left or the right paw was stimulated. The electrical stimulation consisted in square pulses of 0.3 ms at 2 mA, with a frequency of 9 Hz, delivered by an A365 stimulus isolator interfaced with a DS8000 digital stimulator (World Precision Instruments, Stevenage, UK), triggered by the MR scanner. Each fMRI run consisted in blocks of 21 s stimulation and 21 s rest, repeated for five minutes. Consecutive fMRI runs were separated by five minutes of full rest. The same paw was stimulated for two consecutive runs, before switching to the other paw.

The comparison between two types of protocols (i.e. multi-slice bSSFP vs single-slice bSSFP and multi-slice bSSFP vs GE-EPI) was performed using pairs of consecutive runs of stimulating the same paw, randomizing which protocol was acquired first, as detailed in **Table 2**.

All fMRI bSSFP images were acquired with a phase cycling of 180° to center the main frequency on the passband. The bSSFP flip angle was adjusted for each protocol to optimize the extent and flatness of the passband plateau rather than the BOLD amplitude.

For anatomical reference, a high-resolution venogram was acquired using a 3D gradient echo sequence: TE/TR = 15/30 ms / flip angle: 20° / matrix: 256x128x128 / 100-µm isotropic resolution / NA = 2 / TA = 16' (Park et al., 2008).

The data were analyzed using SPM12 (http://www.fil.ion.ucl.ac.uk/spm/software/spm12/). The processing pipeline included slice timing correction, in-plane smoothing (kernel FWHM: 1.5 times the voxel size), and a square response function. Statistical significance was retained at $p < 0.05$ after FWE.

| Protocols compared | Multi-slice vs single-slice bSSFP | GE-EPI vs multi-slice bSSFP |
|---|---|---|
| No. of rats involved | 7 | 7 |
| Total pairs of runs | 14 | 27 |

**Table 2.** Number of rats involved in each comparison and total number of paired runs acquired for a given comparison.

Maximum *t*-score, cluster volume and BOLD signal amplitude were subsequently extracted. BOLD amplitude was estimated from the timecourses of the four voxels with highest *t*-score in the cluster. In a second step, when comparing single-slice to multi-slice bSSFP, the data with higher temporal resolution were downsampled to match the lower temporal resolution and reanalyzed to yield new *t*-scores and cluster volumes.

The SNR for each protocol and each run was estimated using random matrix theory (Veraart et al., 2015). In particular, the image SNR and temporal SNR were compared between the multi-slice and single-slice bSSFP protocols.

Paired *t*-tests were run at the two-sided 5% significance level to determine significantly different metrics between protocols.

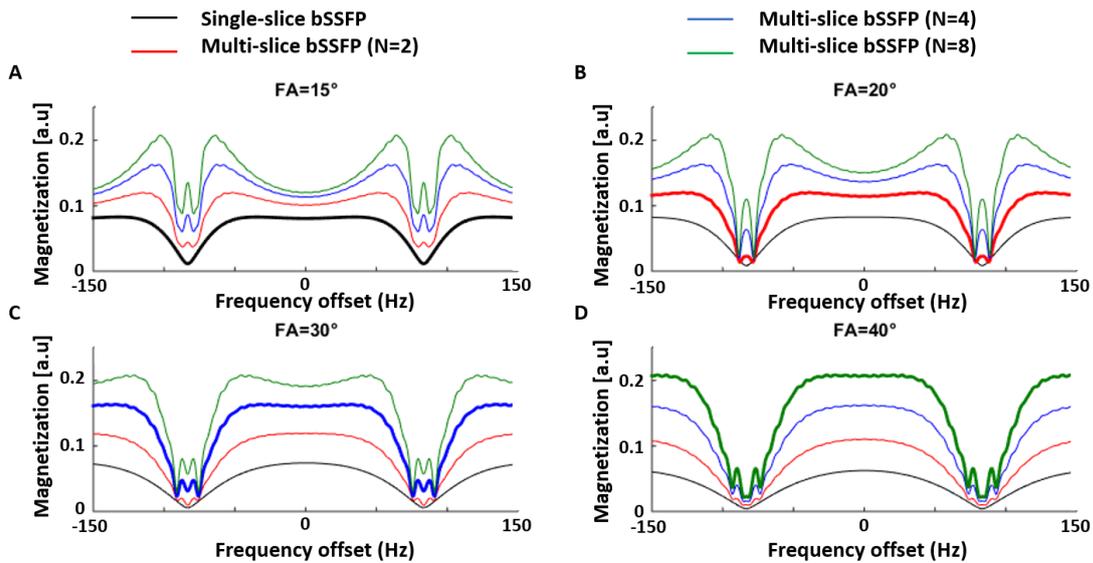

**Figure 2.** Bloch simulation results of bSSFP signal profiles for various flip angles (**A/B/C/D**: FA = 15/20/30/40°) and number of slices (N = 1/2/4/8). All profiles present the pass bands and transition bands characteristic of bSSFP. In multi-slice bSSFP, higher flip angles are required compared to single-slice (true steady-state) in order to maximize the passband plateau. The optimal flip angle is highlighted in bold for N = 1/2/4/8.

## 3. Results
### 3.1. Human brain at 7 Tesla

The evolution of the bSSFP transverse/longitudinal magnetization is illustrated in **Figure 1B**. Compared to conventional (steady-state) bSSFP, higher flip angles were necessary to maximize the multi-slice 2D-bSSFP passband plateau width (**Figure 2**), but the magnetization profiles obtained after FA-adjustment were similar. The 2D-bSSFP pseudo-steady-state benefited from larger longitudinal magnetization recovery during the acquisition of neighboring slices. Simulated bSSFP profile perturbations caused by the phase increment acquisition scheme were minor (**Figure 3A**). The simulations agreed extremely well with the experimental single-slice and multi-slice bSSFP signal profiles (**Figure 3B**). Images confirmed the higher signal level of multi-slice bSSFP (associated with the transient state) (**Figure 3C**). The resting-state networks extracted from single-slice and multi-slice data were similar (in the matching slice), while multi-slice bSSFP benefitted from extended coverage and the identification of networks over a larger spatial scale (**Figure 4**).

Temporal SNR was higher with multi-slice 2D-bSSFP than 3D-bSSFP in GM/WM/WB (**Figure 5A**, *p* < 0.05). Furthermore, 3D-bSSFP suffered from aliasing along the second phase-encode direction because of imperfect slab selection, requiring the outer slices to be discarded (**Figure 5D**). fMRI metrics (number of activated voxels, maximum and mean t-score) were similar between the two schemes (**Figure 5B**). Both acquisition schemes detected a robust activation cluster in the motor cortex during finger-tapping, with no slices to be discarded in the 2D acquisition, as opposed to the 3D one which suffered from fold-over (**Figure 5C-D**).

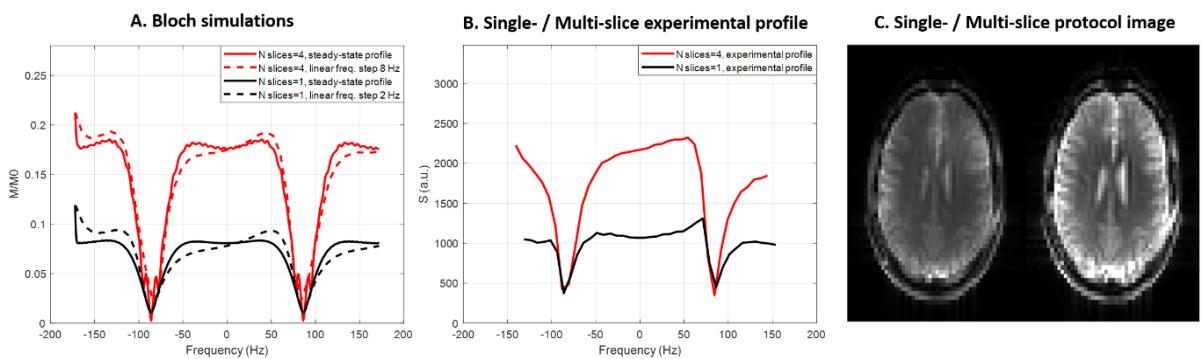

**Figure 3. A**: Simulated signal profiles for passband bSSFP (single-slice/four slices, plain black/red lines, FA = 15/30°) and perturbations due to the linearly increasing phase cycling (dashed lines, equivalent frequency step = 2/8 Hz). **B**: Experimental signal profiles from a single voxel from single slice (black, FA = 15°) and multi-slice bSSFP (red, FA = 30°) obtained by linear increase of the phase cycling. The experimental profiles match the simulations very well. **C**: Signal intensity in single slice (left, FA = 15°, $TR_{vol}$ = 375 ms) and multi-slice bSSFP (right, FA = 30°, $TR_{vol}$ = 1500 ms, one out of four slices displayed). The same scaling was used for both sequences, confirming the higher signal intensity for multi-slice bSSFP.

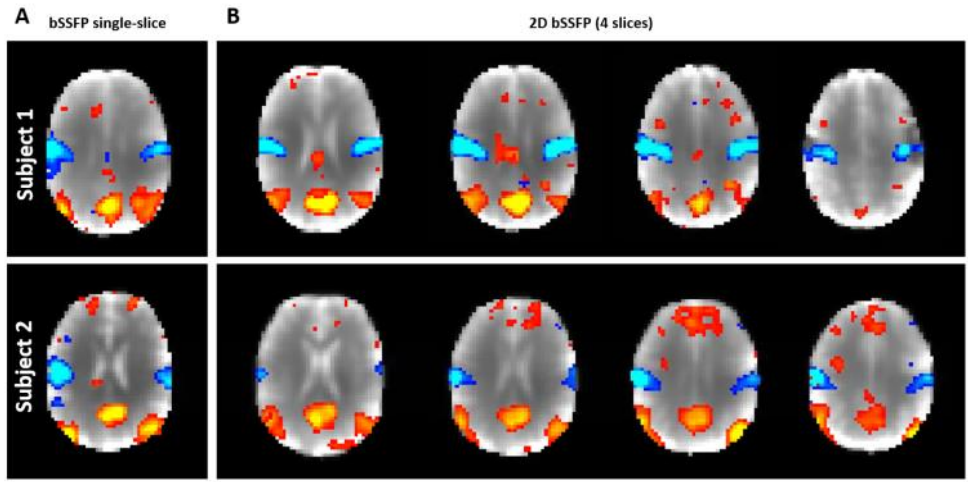

**Figure 4.** Example of two resting-state fMRI data acquired with single-slice and 2D-bSSFP (1/4 slices, TRvol=0.375/1.5s) at 3.5x3.5x4.5 mm resolution (single subject analysis). The Z-score scales in [3-10] for both networks (red – yellow = default mode network, blue – light-blue = sensory-motor). All the RSN detected with single-slice bSSFSP were also detected based on 2D-bSSFP data.

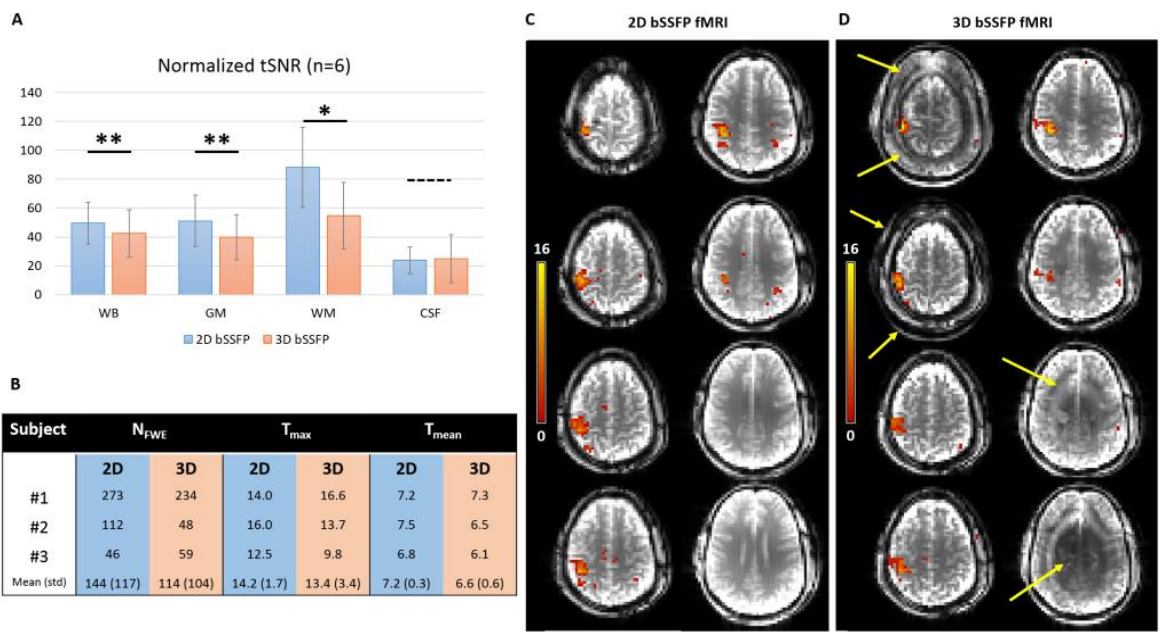

**Figure 5.** 2D vs. 3D-bSSFP fMRI. **A**. A significant tSNR increase was measured for 2D-bSSFP at whole-brain level (WB, +18%, p=0.003), and in mid-brain GM (+29%, p=0.005) and WM (+79%, p=0.014). **B**. Preliminary fMRI results (n=3): 2D and 3D-bSSFP achieved similar performances during finger-tapping (Wilcoxon test, p>0.05). **C-D.** Activation maps (voxelwise 5% FWE threshold) obtained with 2D (**C**) and 3D-bSSFP (**D**) overlaid on the first bSSFP volume for better WM/GM contrast. 3D-bSSFP suffered from aliasing along the transverse direction, contaminating half of the total volume (yellow arrows).

### 3.2. Rat brain at 14 Tesla

The signal during multi-slice bSSFP acquisition at 14T was simulated. While the signal did not reach steady-state during the acquisition of a given slice, the use of 16 dummy scans ensured that the transverse magnetization level was slowly varying (up to 14%) such as not to introduce substantially uneven weighting in *k*-space (**Figure 6**). The optimal flip angle for a passband profile in the steady-state was calculated to be 12° ($\cos \alpha = \frac{T_1/T_2 - 1}{T_1/T_2 + 1}$) at 14 Tesla (Scheffler and Lehnhardt, 2003). In the multi-slice implementation, simulations showed the optimal flip angle to be dependent on sequence parameters (and essentially on the time available for longitudinal relaxation during the acquisition of other slices), with a value of 22° for our protocol, which was also verified experimentally (**Figure 7A**). A few initial volumes needed to be discarded before the slab steady-state was established (**Figure 7B**).

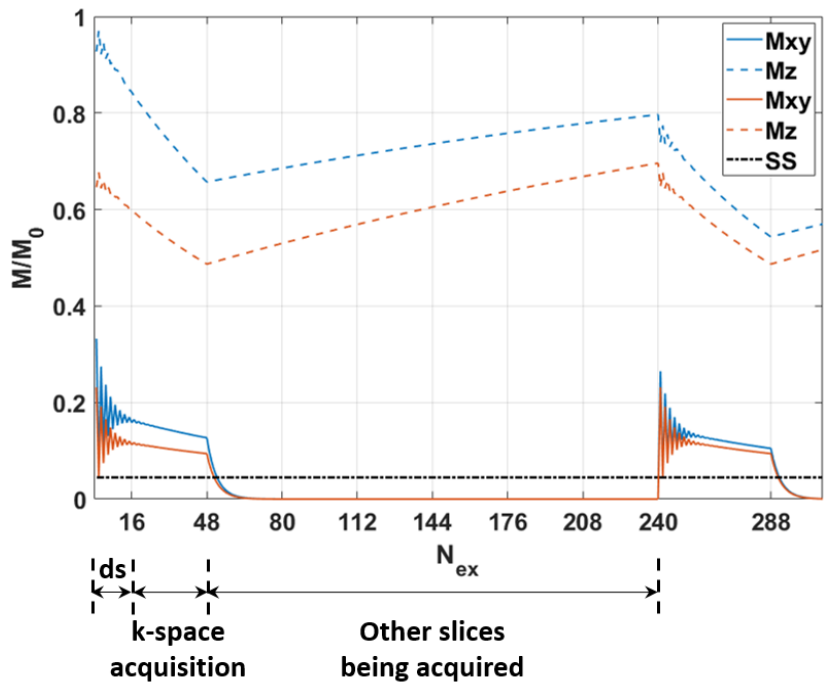

**Figure 6.** Simulated bSSFP signal evolution during the acquisition of five slices, with 16 dummy scans (ds) and 32 phase encoding steps, for TE/TR = 3/6 ms. The blue curves represent the first two consecutive repetitions in a run, where the signal level still varies between repetitions, and the red curves represent two later consecutive repetitions, when slab steady-state is established. Each k-space acquisition still occurs in the transient phase of bSSFP, but the transverse magnetization Mxy is slowly varying. The true steady-state level for Mxy is shown in black.

**Multi-slice bSSFP fMRI**

We compared the multi-slice implementation to the single-slice acquisition at the same TE = 3 ms. Since it exploits the transient state, the SNR of the multi-slice protocol was higher than that of single-slice steady-state: SNR = 58 ± 12 for multi-slice and SNR = 38 ± 7 for single-slice in the somatosensory cortex, as also predicted by simulations (**Figure 7A**). The temporal SNR for the single-slice protocol was naturally higher ($tSNR_{ss}$ = 85 s$^{-1/2}$ vs $tSNR_{ms}$ = 47 s$^{-1/2}$) but the difference in spatial coverage between protocols still speaks in favor of multi-slice bSSFP.

In terms of task-fMRI metrics, both cluster size ($p$ = 4.27E-04) and $t$-score ($p$ = 8.91E-05) were higher with the steady-state sequence vs multi-slice (due to higher tSNR for single slice), though after down-sampling the single-slice to the same temporal resolution, cluster size became similar and *t*-score significantly lower (*p* = 1.35E-04) compared to multi-slice bSSFP (due to the higher image SNR in the latter). The BOLD amplitude was otherwise comparable between the two protocols (1.4 ± 0.4 % and 1.6 ± 0.5 %, respectively) (**Figure 8**).

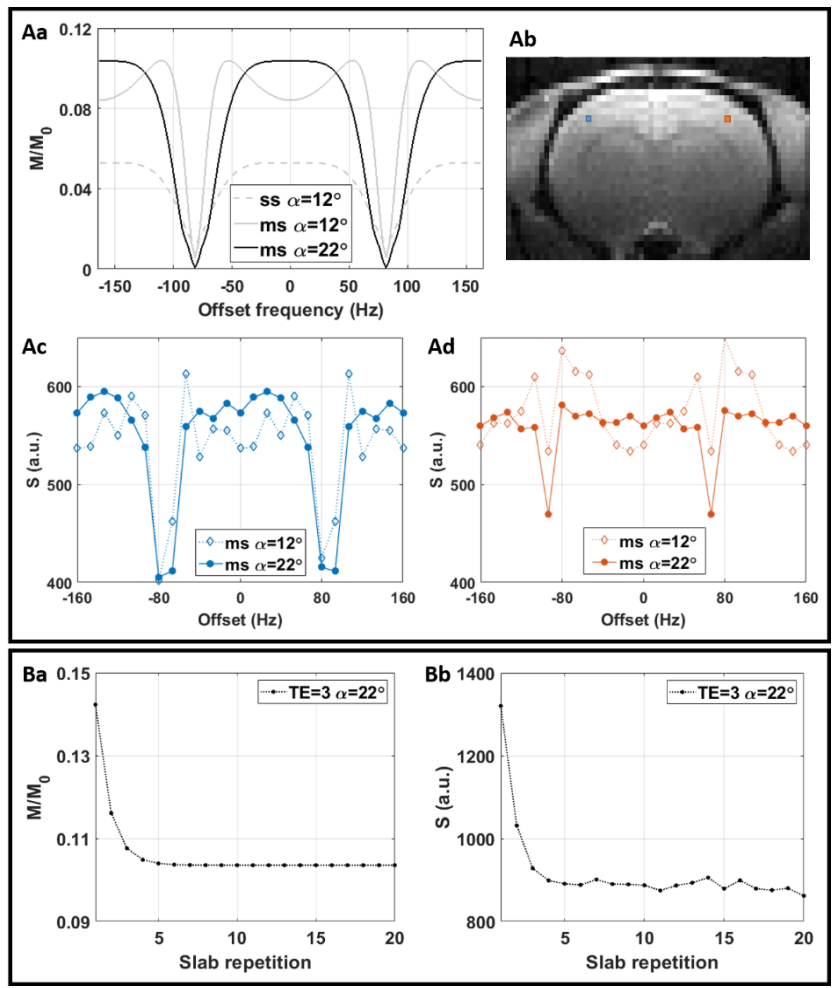

**Figure 7. Panel A**: Single-slice (ss) vs multi-slice (ms) profiles for TE/TR = 3/6 ms. Compared to steady-state signal, the flip angle needed to be increased from 12° to 22° for an optimal passband profile. **Aa**: Simulated signal profile as a function of offset frequency (assuming 180° phase cycling). **Ab:** One voxel in the somatosensory cortex is identified on either side (blue and red). The signal profiles in these two voxels are shown in **Ac** and **Ad**, respectively, for both flip angles. The experimental signal profiles agreed well with the simulations. Note that an experimental profile is a convolution between the theoretical profile and the frequency distribution within the voxel. **Panel B:** Multi-slice bSSFP. Simulations predicted that the slab steady-state was established after six repetitions (**Ba**), which matched the experiment (**Bb**).

Activation clusters were very similar between GE-EPI and multi-slice bSSFP (**Figure 9A**), with bSSFP images additionally more readily registrable to an anatomical reference. While the BOLD amplitude was significantly higher with GE-EPI than bSSFP (*p* = 4.03E-04), cluster size and maximum *t*-score were comparable between the two sequences (**Figure 9B-C**).

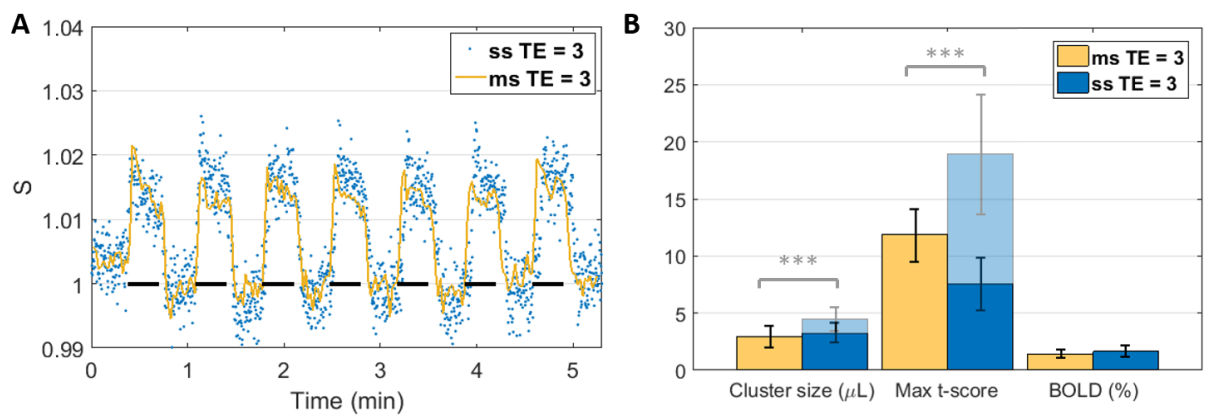

**Figure 8.** Single-slice vs multi-slice protocols with TE = 3 ms. **A**: Mean timecourses, averaged over the four voxels with highest activation from each run. **B**: Comparison of main fMRI metrics. The intrinsic cluster size and maximum t-score for the higher temporal resolution protocol are shown in half-transparency, while the recalculated values at matched temporal resolutions are shown in solid.

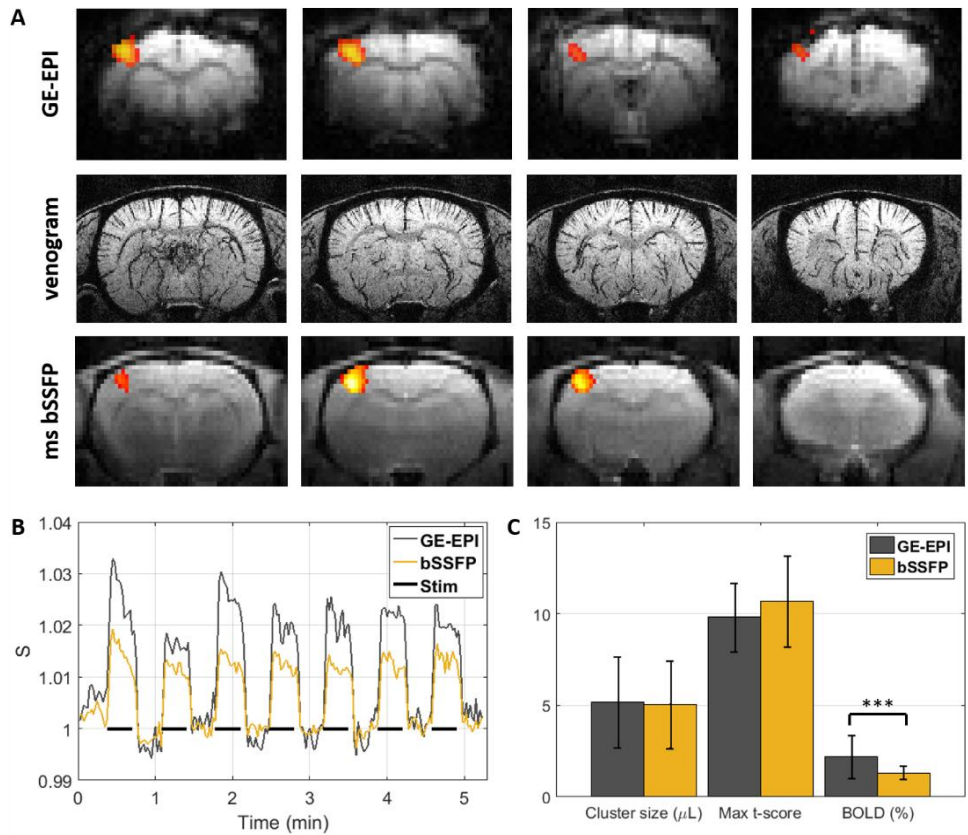

**Figure 9. A:** Example of GE-EPI (top row) and ms bSSFP (bottom row) fMRI results from two consecutive runs with left forepaw stimulation, demonstrating activation of the right somatosensory cortex. T-score maps are overlaid on the functional images (*colorbar limits: 0 – 12*). Middle row: Matching venogram slices. The bSSFP images are more readily registered to the venogram or to any other anatomical reference than GE-EPI. **B**: Mean GE-EPI and ms bSSFP timecourses, averaged over the four voxels with highest activation from each run. **C**: Main fMRI metrics showcasing the differences between the two sequences.

## 4. Discussion and conclusions

This study demonstrates that multi-slice bSSFP is suitable for resting-state and task-fMRI on a thin slab, and provides distortion-free activation maps at high and ultra-high field. A straightforward multi-slice bSSFP implementation with a 180° phase cycling was used, which can be set up on any MR system.

The multi-slice acquisition breaks the steady-state and *k*-space data is acquired in a transient, albeit slowly varying state. The transient state was characterized both through simulations and experimental data, and exhibited similar properties to the steady-state for passband BOLD fMRI purposes, while benefitting from higher signal. The flip angle was increased relative to the optimal steady-state flip angle to retrieve a maximally flat passband. It should be noted that the $T_2$ and $T_1$ weighting of the transient signal are different from that of the steady-state (Scheffler, 2003) which can be expected to impact the BOLD contrast. However, very similar BOLD amplitude and activation maps were obtained between single-slice and multi-slice bSSFP in both human brain (**Figure 4**) and rat brain (**Figure 8**).

On the human 7T system, costs in temporal resolution due to the need to re-establish steady-state between slices (compared to 3D bSSFP) were minimized by the use of catalyzation (Deshpande et al., 2003) to accelerate convergence towards steady-state, and mitigated by higher signal in the new steady-state. 2D-bSSFP tSNR values were higher than 3D, also as a byproduct of reduced motion/physiological noise, as shown for unaccelerated 2D and 3D-GE-EPI (Reynaud et al., 2016), and resulting in superior or equal performances for task-fMRI, in addition to reduced image artefacts. At high field, many regions affected by strong magnetic susceptibility mismatch and that benefit from non-EPI acquisitions – such as the brainstem and retina – are unfortunately also subject to major flow or motion artefacts originating from the blood vessels or the eyes, making 3D-bSSFP acquisitions problematic as well. In that regard, motion-related artifacts are substantially reduced when using a multi-slice approach, and 2D-bSSFP might be better suited than its 3D counterpart for distortion-free localized fMRI applications, such as retina fMRI (Muir and Duong, 2011).

Accelerated 3D-bSSFP has shown great potential for distortion-free imaging at high field applied to fMRI (Chappell et al., 2011; Ehses and Scheffler, 2018; Lee et al., 2008; Scheffler and Ehses, 2016) or Arterial Spin Labeling (Han et al., 2016). Future work could focus on combining 2D-bSSFP with in-slice (GRAPPA) and through-slice (SMS) acceleration, as a less motion-sensitive alternative to accelerated 3D-bSSFP with similar brain coverage and/or spatial resolution. Alternatively, the pseudo steady-state generated by alternating-SSFP with accelerated 3D readout can be optimized as described in this study to generate rapid distortion-free and artefact-free 3D-bSSFP functional contrast at 7T (Reynaud et al., 2018), similarly to a previous implementation at 3T (Jou et al., 2016).

On the rodent 14T system, the multi-slice acquisition represents a compromise between brain coverage (which needs to extend beyond the single slice for proper coverage of a given functional area) and temporal resolution, which would likely be too low in a full Cartesian 3D acquisition. While the availability of multi-channel receive arrays for rodents remains limited, so do acceleration options. One group reported using stacks of spirals for 3D bSSFP fMRI in rats (Lee et al., 2010), while another implemented compressed sensing for single-slice bSSFP fMRI (Han et al., 2015). These acceleration options could in principle be combined with a multi-slice bSSFP acquisition, and a formal comparison between 2D multi-slice and 3D bSSFP fMRI could be performed, but the human data in this work suggest the benefit of 2D multi-slice would be maintained. The pre-clinical implementation of the sequence would also benefit from a ramped preparation scheme to reduce the number of dummy scans required to reach the transient state signal – currently one third of the time was spent on dummy scans. Alternatively, a center-in encoding could be used with fewer dummy scans to produce the same quasi-steady-state for the central k-space line (Jou et al., 2016), but such schemes are more sensitive to eddy currents.

Remarkably, bSSFP provided similar activation maps to GE-EPI in the rodent brain. In spite of a lower BOLD amplitude for bSSFP, related to the shorter TE than that of GE-EPI and the different contrast mechanisms, activation epochs were detected very reliably, achieving similar cluster size and *t*-statistics. We underline that our goal was to perform a practical comparison between the two sequences by matching their temporal and spatial resolutions, as well as brain coverage, rather than a formal comparison of BOLD contrast mechanisms between bSSFP and GE (e.g. by matching their TE's) as investigated by other groups (Miller et al., 2007; Park et al., 2011; Zhong et al., 2007). Previous practical comparisons of bSSFP to GE-EPI performed at 7 T in rats (Cheng et al., 2014; Muir and Duong, 2011), indicating similar activation detection for both task and resting-state fMRI, seem to extend to 14 T. It should be noted that bSSFP images are also less prone to distortions by field inhomogeneity, very prominent at ultra-high field, and are more easily registered to an anatomical reference than GE-EPI. The multi-slice bSSFP acquisition is expected to be increasingly beneficial for mouse fMRI: the smaller skull size of mice leads to larger distortions in GE-EPI images, but is not expected to affect bSSFP images in a major way.

## Acknowledgments

The authors thank Klaus Scheffler and Valerij Kiselev for insightful discussions. They also acknowledge assistance with animal setup and monitoring from Mario Lepore and Stefan Mitrea. This work was supported by the Centre d'Imagerie Bio-Médicale (CIBM) of the University of Lausanne (UNIL), the Swiss Federal Institute of Technology Lausanne (EPFL), the University of Geneva (UniGe), the Centre Hospitalier Universitaire Vaudois (CHUV), the Hôpitaux Universitaires de Genève (HUG) and the Leenaards and the Jeantet Foundations.